\documentclass[aps,prd,preprint,groupedaddress]{revtex4-1}

\usepackage{latexsym}
\usepackage{amsthm}
\usepackage{amsmath}
\usepackage{graphicx}
\usepackage{mathtools}
\usepackage{slashed}
\usepackage{amssymb}
\usepackage{indentfirst}
\usepackage{subfigure}
\usepackage{physics}
\usepackage{color}

\newcommand*\diff{\mathop{}\!\mathrm{d}}
\newcommand{\nn}{\nonumber}

\newcommand{\be}{\begin{eqnarray}}
\newcommand{\ee}{\end{eqnarray}}
\newcommand{\ma}{\mathrm}
\newcommand{\ml}{\mathcal}
\newcommand{\bs}{\boldsymbol}

\begin{document}

\title{Dynamical Screening of $\alpha$-$\alpha$ Resonant Scattering and Thermal Nuclear Scattering Rate in a Plasma}

\author{Xiaojun Yao}
	\email{xiaojun.yao@duke.edu}
\author{Thomas Mehen}
\author{Berndt M\"uller}
\affiliation{Department of Physics, Duke University, Durham, NC 27708, USA}

\date{\today}

\begin{abstract}
We use effective field theory and thermal field theory to study the dynamical screening effect in the QED plasma on the $\alpha$-$\alpha$ scattering at the $^8$Be resonance. Dynamical screening leads to an imaginary part of the potential which  results in a thermal width for the resonance and dominates over the previously considered static screening effect. As a result, both the resonance energy and width increase with the plasma temperature. Furthermore, dynamical screening can have a huge impact on the $\alpha$-$\alpha$ thermal nuclear scattering rate. For example, when the temperature is around $10$ keV, the rate is suppressed by a factor of about $900$. We expect similar thermal suppressions of nuclear reaction rates to occur in those reactions dominated by an above threshold resonance with a thermal energy. Dynamical screening effects on nuclear reactions can be relevant to cosmology and astrophysics.
\end{abstract}

\maketitle

\section{Introduction}
Within a QED plasma consisting of electrons and positrons, the photon effectively gains a mass called the Debye mass, $m_D$, that screens the Coulomb interaction by an exponential factor, $e^{-m_Dr}$, which is known as static screening. Its effect on the $\alpha$-$\alpha$ resonant scattering, where the resonant state is the $^8$Be nucleus, was recently studied by applying effective field theory (EFT) and thermal field theory \cite{Yao:2016gny}. The $^8$Be nucleus has isospin $I=0$ and lies at the center-of-mass (CM) energy $E_{0}=91.84\pm0.04$ keV with a width $\Gamma_0=5.57\pm0.25$ eV in vacuum. Inside the QED plasma it was found that the energy of the $^8$Be resonance decreases and its lifetime increases with the plasma temperature \cite{Yao:2016gny}.

However, static screening is not the only plasma effect that modifies low energy $\alpha$-$\alpha$ scattering. In addition, the medium particles constantly collide with the $\alpha$ particle and change its momentum. This leads to an imaginary part of the $\alpha$-$\alpha$ interaction potential, which accounts for the Landau damping rate for an $\alpha$-$\alpha$ state with given momenta. This is called dynamical screening. The imaginary part of the potential in a plasma has been derived in Ref.~\cite{Laine:2006ns,Brambilla:2008cx} for the quark-antiquark color interaction and in Ref.~\cite{Beraudo:2007ky} for the electric interaction.  Later, the resultant complex potential was used in  phenomenological studies of quarkonia spectral functions \cite{Petreczky:2010tk} and quarkonia dynamics \cite{Brambilla:2013dpa, Blaizot:2015hya, Burnier:2015tda} in the quark-gluon plasma (QGP), which is assumed to be produced in relativistic heavy ion collisions.

The imaginary potential can also be interpreted in terms of the open quantum system formalism, where a system is coupled with a thermal medium and they evolve as a whole in time. When the medium part is traced out, the system evolves non-unitarily. The non-unitary evolution can be related  to the influence of an imaginary potential. In this way the medium effect on the time evolution of the system can be studied more generally. This idea has been pursued to study the quarkonium evolution in the QGP \cite{Young:2010jq,Borghini:2011ms,Akamatsu:2011se}, where it was shown how the spatial decoherence of the quarkonium wave function leads to the suppression of the quarkonium state in the QGP \cite{Akamatsu:2011se}. 

Inspired by the studies of dynamical screening effects on quarkonia, we study their effects on $\alpha$-$\alpha$ resonant scattering. We show that the dynamical screening effects are much larger than the static ones, and as a result, the energy and width of the $^8$Be resonance increase with the plasma temperature. We also compute the thermally averaged $\alpha$-$\alpha$ nuclear scattering rate and find that at temperatures just below the vacuum resonance energy, dynamical screening can drastically change the thermal nuclear scattering rate. The effect is most significant in the range $1\,{\rm keV} < T < 75\,{\rm keV}$. These temperatures are relevant to both cosmology, especially Big Bang Nucleosynthesis, and stellar astrophysics. 
The framework developed here for including dynamical screening effects in thermal rates can be applied to nuclear reactions relevant to cosmology and astrophysics.  

The article is organized as follows. In Sec.~\ref{sect:eft} the EFT describing the $\alpha$-$\alpha$ interaction is briefly reviewed and the scattering amplitude is derived. In Sec.~\ref{sect:imag} the imaginary part of the interaction potential is derived using the thermal photon propagator. By including the imaginary potential, the modified resonance energy and width are calculated in Sec.~\ref{sect:width}. Then the dynamical screening effect on the $\alpha$-$\alpha$ thermal nuclear scattering rate is discussed in Sec.~\ref{sect:rate}. Finally,  conclusions are drawn in Sec.~\ref{sect:concl}.

\section{Effective Lagrangian and Scattering Amplitude}
\label{sect:eft}
The effective Lagrangian for low-energy $\alpha$-$\alpha$ scattering is \cite{Kaplan:1996xu,Kaplan:1998we,Kong:1999sf,Higa:2008dn} 
\be
\ml{L}=N^{\dagger}\bigg(iD_t+\frac{{\bs D}^2}{2M}\bigg)N -  \frac{C_0}{4}N^{\dagger}N^{\dagger} N N + \frac{C_2}{32}\Big(N^{\dagger} \overleftrightarrow{\nabla}^2 N^\dagger N  N  + h.c. \Big) +\cdots  ,
\ee
where the field $N$ represents the $\ma{\alpha}$ particle and $M\approx3727.38$ MeV is its mass. The four-point vertex with an incoming momentum ${\bs p}$ in the CM frame is
$-i\sum_{n=0}^{\infty}C_{2n}p^{2n}$. The propagator of a single $\alpha$ particle is $\frac{i}{E-{\bs p}^2/(2M)+i\epsilon}$.
The scattering amplitude $T({\bs p'}, {\bs p})$ can be written as two parts: one is the pure Coulomb scattering amplitude $T_C({\bs p}', {\bs p})$ and the other the Coulomb-modified strong scattering amplitude $T_{SC}({\bs p'}, {\bs p})$. The latter has a Lippmann-Schwinger expansion
\be
\label{eq:scatt_ampsc}
T_{SC}({\bs p'}, {\bs p}) = \langle \psi_{\bs p'}^{(-)} | V_S | \Psi_{\bs p}^{(+)} \rangle
= \sum_{n=0}^{\infty}\langle \psi_{\bs p'}^{(-)} | V_S \big(\hat{G}_C^{(+)}(E)V_S \big)^n| \psi_{\bs p}^{(+)} \rangle \,,
\ee
where the scattering in-state $(+)$ and out-state $(-)$ are defined as $(H_0+V_C) | \psi_{\bs p}^{(\pm)} \rangle = E| \psi_{\bs p}^{(\pm)} \rangle$ and
$(H_0+V_C +V_S) | \Psi_{\bs p}^{(\pm)} \rangle = E| \Psi_{\bs p}^{(\pm)} \rangle$. Here, $V_C$ is the Coulomb potential (in vacuum, $V_C=Z_1Z_2\alpha/r$) and $V_S$ is the strong interaction potential. (For $\alpha$-$\alpha$, $Z_1=Z_2=2$, but we will leave $Z_1$ and $Z_2$ arbitrary so our results can be applied to more general reactions.) Inserting a complete set of position eigenstates and using the fact that $V_S$ is a delta function in position space (with a coefficient depending on the energy) leads to
\be
\label{Tmat}
T_{SC}({\bs p'}, {\bs p})&=&-\frac{\psi_{\bs p'}^{(-)*}(0)\psi_{\bs p}^{(+)}(0)}{(\sum_nC_{2n}p^{2n})^{-1}-G(E,0,0)} \, ,
\ee
where the negative sign is a convention in quantum field theory Feynman diagram calculations. The Coulomb wave function can be expressed as
\be
\psi_{\bs p}^{(+)}({\bs r}) &=& e^{-\pi\eta/2}\Gamma(1+i\eta)M(-i\eta,1;ipr-i{\bs p}\cdot{\bs r})e^{i{\bs p}\cdot{\bs r}} \nn\\
\psi_{\bs p}^{(-)} ({\bs r}) &=& e^{-\pi\eta/2}\Gamma(1-i\eta)M(i\eta,1;-ipr-i{\bs p}\cdot{\bs r})e^{i{\bs p}\cdot{\bs r}} \, ,
\ee
where $M(a,b;z)$ is the confluent hypergeometric function and $\eta=Z_1Z_2\alpha M/2p$. The Sommerfeld factor is defined as $|\psi_{\bs p}^{(\pm)}(0)|^2=C_{\eta}^2=\frac{2\pi\eta}{e^{2\pi\eta}-1}$. Then the numerator $\psi_{\bs p'}^{(-)*}(0)\psi_{\bs p}^{(+)}(0)$ with the elastic condition $|{\bs p'}|=|{\bs p}|$ is equal to $C_{\eta}^2e^{2i\sigma_0}$ where $\sigma_0=\arg{\Gamma(1+i\eta)}$ is the phase shift caused by the Coulomb interaction only.
The retarded Coulomb Green's function in the spatial representation is given by
\be
G(E,0,0)&=&\Big\langle {\bs r}'=0 \Big|\hat{G}_C^{(+)}(E) \Big|{\bs r}=0 \Big\rangle = \bigg\langle 0 \bigg|   \frac{1}{E-\hat{H}_0-V_C + i\epsilon}    \bigg|    0 \bigg\rangle \nn \\
&=& \frac{Z_1Z_2\alpha M^2}{4\pi}\bigg( \frac{1}{\epsilon}-H(\eta)+\ln{\frac{\mu\sqrt{\pi}}{Z_1Z_2\alpha M}}+1-\frac{3}{2}\gamma \bigg) \,,
\ee
where the second line is the expression in the MS renormalization scheme with scale $\mu$ \cite{Kong:1999sf}. Here, $\gamma$ is the Euler constant and $H(\eta)=\psi(i\eta)+\frac{1}{2i\eta}-\ln{(i\eta)}$, where $\psi(z)$ is the digamma function.
Expanding the first term in the denominator of Eq.~(\ref{Tmat}) to order  $p^4$ (which preserves unitarity)  leads to
\be
T_{SC}&=&\frac{C_{\eta}^2e^{2i\sigma_0}}{-\frac{1}{C_0} +\frac{C_2}{C^2_0}p^2-\Big( \frac{C_2^2}{C_0^3}-\frac{C_4}{C_0^2} \Big)p^4+G(E,0,0)} \, .
\ee
We follow the conventions and renormalization scheme in Ref.~\cite{Yao:2016gny}, where the divergent and energy-independent terms in $G(E,0,0)$ are absorbed into the 
definition of $C_0$, and obtain the following expression for  $T_{SC}$:
\be
T_{SC}&=&\frac{4\pi}{M}\frac{C_{\eta}^2e^{2i\sigma_0}}{-\frac{1}{a} + \frac{r_0}{2} p^2 - \frac{P_0}{4} p^4-Z_1Z_2\alpha MH(\eta)} \, ,
\ee
where $T_{SC}$ is expressed in terms of the effective range expansion parameters: the scattering length $a$, the effective range $r_0$ and the shape parameter $P_0$. The relationship between the $C_i$ and the effective range expansion parameters can be found in Ref.~\cite{Yao:2016gny}, which also fitted the parameters to reproduce the resonance properties and measured S-wave phase shift, which dominates in the low-energy scattering, up to $E_{CM}=3$ MeV. (A similar fit was performed 
in Ref.~\cite{Higa:2008dn}.) The result of the  fit is shown in Table~\ref{table:1}, these parameters will be used in our calculations for the remainder of this paper.
\begin{table*}
\caption{\label{table:1}Best fit parameters from Ref.~\cite{Yao:2016gny}}
\begin{center}
\begin{tabular}{|c|c|c|c|}
\hline
Parameter & $a\ (10^3\ \ma{fm})$ & $r_0\ (\ma{fm})$ & $P_0\ (\ma{fm}^3)$\\
\hline
Best fit value (accurate to $10^{-3}$) &  -2.029 & 1.104 &  -1.824  \\
\hline
\end{tabular}
\end{center}
\end{table*}

\section{Imaginary Part of Potential}
\label{sect:imag}

\begin{figure}
\centering
\includegraphics[width=4.5in]{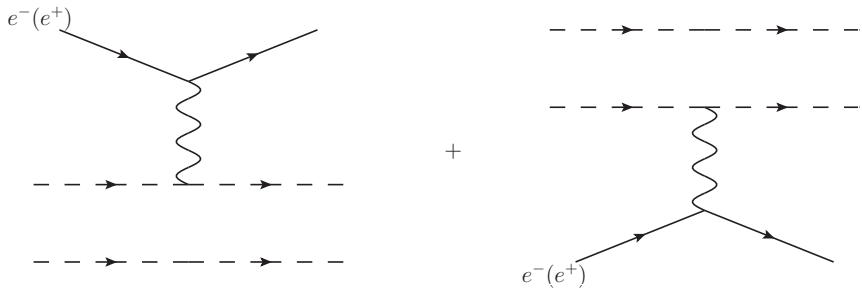}
\caption{Leading order Feynman diagrams contributing to the damping rate of an $\alpha$-pair. The dashed line indicates an $\alpha$ particle. 
}
\label{fig:damping}
\end{figure}
\begin{figure}
\centering
\includegraphics[width=6.0in]{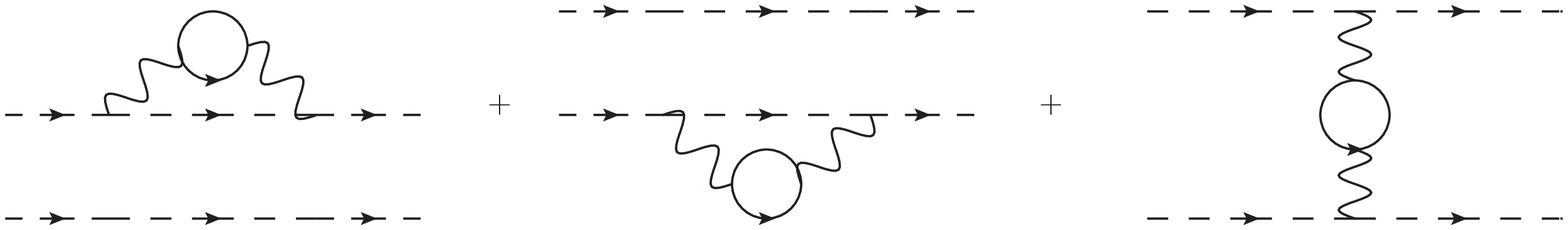}
\caption{Loop corrected $\alpha$-$\alpha$ forward scattering amplitude, which contains the lowest-order contribution to the imaginary potential. The solid line in loops indicates electron or positron. These diagrams can also represent the loop corrections to single $\alpha$ propagators and the Coulomb exchange interaction.}
\label{fig:imaginary}
\end{figure}

Generally, when a charged particle moves in a QED plasma, its momentum will no longer be a constant because the medium particles (electrons or positrons) constantly scatter with it.  This elastic scattering can change the relative momentum of an $\alpha$ pair but not their total kinetic energy. This leads to an imaginary part in the potential, which describes the rate for losing a charged particle state with a given momentum in the plasma, a phenomenon known as Landau damping. It can be calculated at leading order (LO) from Fig.~\ref{fig:damping} by taking the square of the scattering amplitude, summing over the final state of the $\alpha$ particles and averaging thermally over the medium particles. According to unitarity and the cutting rules, this corresponds to the imaginary part of the $\alpha$-$\alpha$ forward scattering amplitude, shown in Fig.~\ref{fig:imaginary}. We will extract the imaginary potential from these diagrams and then compute the $\alpha$-$\alpha$ particle Green's function including this imaginary potential.  Fig. 2 also shows the thermal loop corrections to the single $\alpha$ propagators and the Coulomb exchange interaction. 
Both corrections contribute to the Coulomb potential, which is long-ranged and sensitive to the Debye mass because it is the typical momentum transferred between medium particles and the $\alpha$ particle. Therefore, when studying the thermal loop corrections of the Coulomb potential, one expects the loop momentum $|\bs q|\sim m_D$ and the loop energy $q_0\sim \bs q^2/M\ll m_D$. For temperatures $T<1$ MeV, the hierarchy of scales $q_0\ll |\bs q| \ll T,m_e$ is valid and one can make the hard thermal loop (HTL) approximation including a finite electron mass $m_e$. The time-ordered thermal photon propagator under this approximation has been calculated in Ref.~\cite{Escobedo:2008sy}:
\be
D_{00}(q_0=0,\bs{q})=\frac{i}{\bs{q}^2+m_D^2} + \frac{16\alpha g(m_e\beta)}{|\bs{q}|(\bs{q}^2+m_D^2)^2\beta^3} \, .
\ee
The error on the imaginary potential introduced by this approximation is discussed in the Appendix. The Debye mass is given by \cite{Escobedo:2008sy}
\be
m_D^2 &=&\frac{8m_e^2}{(2\pi)^2}e^2(2f(m_e\beta)+h(m_e\beta)) \, ,
\ee
where the functions $f$, $g$ and $h$ are defined as
\be
f(m_e\beta) &=& \frac{1}{m_e^2}\int_0^{\infty} \diff k \frac{k^2}{\sqrt{k^2+m_e^2}(e^{\beta\sqrt{k^2+m_e^2}}+1)}=-\sum_{n=1}^{\infty}(-1)^n\frac{K_1(n\beta m_e)}{n\beta m_e}\\
h(m_e\beta) &=& \int_0^{\infty} \diff k \frac{1}{\sqrt{k^2+m_e^2}(e^{\beta\sqrt{k^2+m_e^2}}+1)}=-\sum_{n=1}^{\infty}(-1)^nK_0(n\beta m_e)\\
g(m_e\beta) &=& \beta^2\int_0^{\infty} \diff k \frac{k}{e^{\beta\sqrt{k^2+m_e^2}}+1} = m_e\beta\ln{(1+e^{-m_e\beta})} -Li_2(-e^{-m_e\beta}) \, ,
\ee
where $K_0(x)$ and $K_1(x)$ are the modified Bessel functions and $Li_2(x)$ is the dilogarithmic function. For low temperatures $m_e\beta \gg 1$,
these functions are approximated by
\be
m_D^2  &=& 8\alpha \sqrt{\frac{m_e^3}{2\pi\beta}}e^{-m_e\beta}\left[1+ \ml{O} \left(\frac{1}{m_e\beta}\right) \right],\\
g(m_e\beta) &=&  (m_e\beta +1) e^{-m_e\beta} +\ml{O}(m_e \beta e^{-2m_e \beta}) \, .
\ee
In the limit $m_e \beta\to 0$ we recover the standard HTL result with massless electrons, $m_D^2 = 4 \pi \alpha T^2/3$, $g(0) = \pi^2/12$, and the second term in $D_{00}(q_0=0,\bs{q})$ becomes $\frac{\pi m_D^2T}{q(q^2+m_D^2)^2}$.

In the infinite $\alpha$ particle mass approximation, we neglect the kinetic energy term in the $\alpha$ particle propagator. Then each of the first two diagrams in Fig.~\ref{fig:imaginary} contributes to the $\alpha$ particle (time-ordered) self energy
\be
\label{eqn:sigma}
i\Sigma_{1(2)} &=& (iZ_{1(2)}e)^2\int \frac{\diff^4q}{(2\pi)^4}\frac{i}{q_0+i\epsilon}D_{00}(q_0,{\bs q})  \nn \\
&=& i(iZ_{1(2)}e)^2\int \frac{\diff^4q}{(2\pi)^4} \bigg[\ml{P}\frac{1}{q_0}-i\pi\delta(q_0)\bigg]D_{00}(q_0,\bs{q}) \nn \\
&=&
-\frac{1}{2}(Z_{1(2)}e)^2\int \frac{\diff^3q}{(2\pi)^3} D_{00}(q_0=0,\bs{q})= iZ_{1(2)}^2\bigg(\frac{1}{2}\alpha m_D + i\frac{8\alpha^2g(m_e\beta)T^3}{\pi m_D^2}   \bigg) \, ,
\ee
where in the second line the principle value vanishes because $D_{00}(q_0,\bs{q})=D_{00}(-q_0,\bs{q})$ and the integrand is odd in $q_0$. In the last line a linear divergence that survives in the zero-temperature limit has been absorbed into a renormalization of the $\alpha$ particle mass. Then the single $\alpha$ particle propagator becomes
\be
\frac{i}{E-\frac{{\bs p}^2}{2M}+i\epsilon + \Sigma_{1(2)}}= \frac{i}{E-\frac{{\bs p}^2}{2M}+i\epsilon +Z_{1(2)}^2\big(\frac{1}{2}\alpha m_D + i\frac{8\alpha^2g(m_e\beta)T^3}{\pi m_D^2}   \big)} \, .
\ee
The third diagram modifies the Coulomb exchange potential 
\be
\label{eqn:coulomb}
V_C({\bs r}) &=& i(iZ_1e)(iZ_2e)\int\frac{\diff^3q}{(2\pi)^3}e^{i{\bs q}\cdot{\bs r}} D_{00}(q_0=0,\bs{q})
\nn \\
&=& \frac{Z_1Z_2\alpha}{ r}e^{-m_Dr} -iZ_1Z_2e^2\int\frac{\diff^3q}{(2\pi)^3}e^{i{\bs q}\cdot{\bs r}}\frac{16\alpha g(m_e\beta)T^3}{q(q^2+m_D^2)^2} \, ,
\ee
where the first term is the static screened Coulomb potential while the second term is the dynamical screening contribution. 

\begin{figure}
\centering
\includegraphics[width=3in]{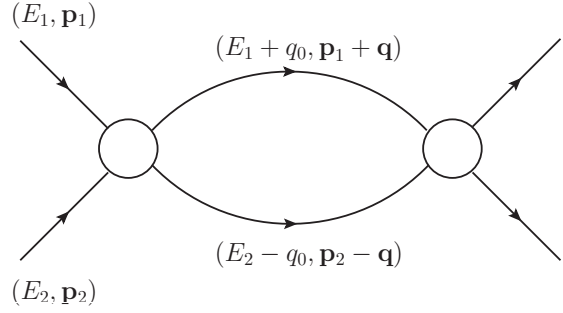}
\caption{A loop correction to the two-$\alpha$ propagator for arbitrary energy and momentum. The white circle can be either a strong interaction vertex or a screened Coulomb exchange. }
\label{fig:loop_rel}
\end{figure}

The damping rate comes from collisions with medium particles as in Fig.~\ref{fig:damping} which  change both the CM and relative momenta of the $\alpha$ particle pair. However, only the effect on the relative motion is relevant for the resonance properties. One needs to figure out how $\Sigma_{1(2)}$ modifies the Green's function associated with the Hamiltonian of the relative motion. This can be seen as follows: consider the one-loop correction to the 
two-$\alpha$ propagator  in Fig.~\ref{fig:loop_rel}, where the energies and momenta of incoming particles are labelled in an arbitrary reference frame, which is proportional to 
\be
\label{eqn:free_prop}
&&\int\frac{\diff^4q}{(2\pi)^4}\frac{i}{E_1+q_0- (\bs p_1 + \bs q)^2/2M +\Sigma_1 +i\epsilon}   \frac{i}{E_2-q_0-(\bs p_2 - \bs q)^2/2M + \Sigma_2+i\epsilon} \, \nn \\
&=&  i \int\frac{\diff^3q}{(2\pi)^3} \frac{1}{E_1 + E_2 -(\bs p_1+\bs q)^2/2M  - (\bs p_2 - \bs q)^2/2M + \Sigma_1 + \Sigma_2+i\epsilon}\nn \\
&=&i \int\frac{\diff^{3}q}{(2\pi)^{3}}\frac{1}{E_1 +E_2 - ({\bs p_1 +\bs p_2)}^2/(4M) -{\bs q}^2/M+\Sigma_1+\Sigma_2 +i\epsilon} \nn \\ 
&=& \int\frac{\diff^{3}q}{(2\pi)^{3}}\frac{i}{E_{CM} -{\bs q}^2/M+\Sigma_1+\Sigma_2+i\epsilon} 
\, .
\ee
Contributions in Fig.~\ref{fig:loop_rel} from the white circles that can be either a strong interaction vertex or a screened Coulomb exchange, both of which are independent of $q_0$ (the $q_0$-dependent part in the Coulomb case is suppressed by the large mass factor),  are omitted in this expression. The second line follows from a contour integral and the third line from a shift in $\bs q$ which is allowed because the integration is over all $\bs q$. 
The combination of $E_i$ and $p_i$ appearing in the denominator of the third line is the Galilean invariant combination that corresponds to the energy of the two $\alpha$ particles in their center of mass, $E_{CM}$. In what follows we will drop the subscript in $E_{CM}$ with the understanding that this is the relevant energy for the Green's function. Eq.~(\ref{eqn:free_prop}) also shows that the self-energy correction of each single $\alpha$ particle propagator enters the Green's function as a sum.

Let $\hat{H}_0$ represent the kinetic energy operator for the relative motion. Then the Green's function for the relative motion between two free $\alpha$ particles, including their individual widths, is 
\be
\hat{G}^+_0(E) = \frac{1}{E-\hat{H}_0 + \Sigma_1 + \Sigma_2+i\epsilon} \,.
\ee
The Coulomb Green's function is given by the Lippmann-Schwinger series
\be
\hat{G}^+_C(E)&=&\hat{G}^+_0(E)+\hat{G}^+_0(E)V_C\hat{G}^+_0(E)+\hat{G}^+_0(E)V_C\hat{G}^+_0(E)V_C\hat{G}^+_0(E)+\cdots \nn\\
&=&\frac{1}{E-\hat{H}_0 + \Sigma_1 + \Sigma_2-V_C+i\epsilon} \nn \\
&=&\frac{1}{E-\hat{H}_0-\frac{Z_1 Z_2\alpha}{r} e^{-m_Dr} + \frac{1}{2}( Z_1^2+Z_2^2)\alpha m_D + iW({\bs r})   +i\epsilon} \, ,
\ee
where
\be\label{Wr}
W(r) &=& e^2\int\frac{\diff^3q}{(2\pi)^3}\Big(\frac{1}{2}( Z_1^2+Z_2^2)+ Z_1 Z_2e^{i{\bs q}\cdot{\bs r}}\Big)\frac{16\alpha g(m_e\beta)T^3}{q(q^2+m_D^2)^2} \nn \\
&=&\frac{16\alpha^2g(m_e\beta)T^3}{\pi m_D^2}\phi(m_Dr, Z_1,Z_2) \, ,
\ee
and
\be\label{phir}
\phi(m_Dr,Z_1,Z_2) = 2\int_0^{\infty}\frac{x\diff x}{(1+x^2)^2}\bigg(\frac{1}{2}(Z_1^2+Z_2^2) + Z_1 Z_2\frac{\sin{(xm_Dr)}}{xm_Dr} \bigg) \, .
\ee

First it is interesting to consider the behavior of the potentials in the limit $ r\to 0$. Except for the unscreened Coulomb interaction $Z_1Z_2\alpha/r$, the real contribution from the self-energies combines with the contribution 
from the static screening of the Coulomb potential to give a negative shift of the potential of $(Z_1+Z_2)^2 \alpha m_D/2$. It is also easy to see from Eqs.~(\ref{Wr},\ref{phir}) that 
$W(0) \propto (Z_1+Z_2)^2$. This shows that both potentials vanish at the origin when the two particles have equal and opposite charges. Two oppositely charged particles 
placed at the same point  appear to the plasma like a neutral particle, in which case the plasma will have no effect on their energy.

\begin{figure}
\centering
\includegraphics[width=3in]{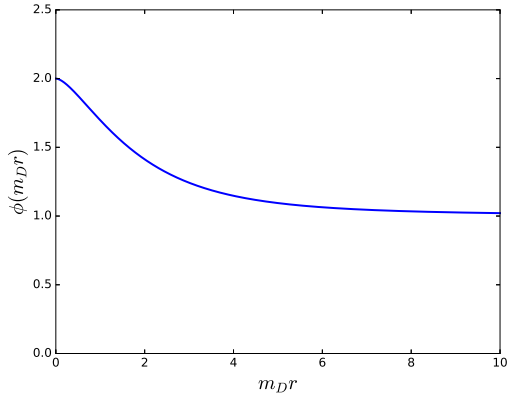}
\caption{The $r$-dependence part of the imaginary potential $\phi(m_Dr)$.}
\label{fig:phi}
\end{figure}

Henceforth we restrict ourselves to the case $Z_1= Z_2 = Z$, then $\phi(m_Dr,Z,Z) = Z^2 \phi(m_D r)$ and
the function $\phi(m_Dr)$ is plotted in Fig.~\ref{fig:phi}. It can be seen that $\phi(0)=2$ and $\phi(\infty)=1$. When the two 
$\alpha$ particles are far separated, the total damping rate is the sum of the individual damping rate of each $\alpha$ particle while when they are close, the damping rate is doubled due
 to their interactions. 

Finally we would like to understand the relative importance of static versus dynamical screening effects. Ref.~\cite{Yao:2016gny} emphasized that for static screening the energy shift of the resonance is to a good approximation linear in the Debye mass, and to a good approximation the temperature-dependent corrections can be obtained by expanding the screened potential to lowest order in $m_D$ which simply adds a constant to the unscreened Coulomb potential of $-Z^2 \alpha \,m_D$. When $T\ll m_e$, the real static screening correction is $Z^2 \alpha \,m_D \sim Z^2 \alpha^{3/2} (m_e^3 T)^{1/4} e^{-m_e/2T}$. In the same limit,
the coefficient of $\phi(m_Dr)$ in the imaginary part of the potential scales as $ Z^2\alpha\sqrt{\frac{T^3}{m_e}}$.    We see that the static screening is suppressed relative to dynamical screening by $\alpha^{1/2}(m_e/T)^{5/4} e^{-m_e/2T}$ for  $m_e \gg T$.  In the opposite limit, $m_e \to 0$, the coefficient of $\phi(m_Dr)$ is $Z^2 \alpha T$, while static screening correction to the potential  is $Z^2\alpha m_D \sim Z^2 \alpha^{3/2} T$, so static screening is suppressed relative to dynamical screening by a factor of $\sqrt{\alpha}$. Thus, in either limit dynamical screening should be expected to be more important.  In our calculations below, which are restricted to $T \lesssim m_e$, dynamical screening dominates.

\section{Thermal Width}
\label{sect:width}
The renormalized Coulomb Green's function is known analytically in vacuum,
\be
\frac{4\pi}{M}G_{\rm ren}(E,0,0;T=0)=-Z^2\alpha MH(\eta).
\ee
In the plasma, 
\be
G(E,0,0;T\neq0) &=& \Big\langle {\bs r'}={\bs 0} \Big|  \frac{1}{E-\hat{H}_0-\frac{Z^2\alpha}{r}e^{-m_Dr}+ Z^2\alpha m_D + iW(r) +i\epsilon}  \Big|  {\bs r}''={\bs 0}  \Big\rangle.
\ee
The complex potential induced by the plasma screening depends on the dimensionless combination $m_Dr$. The characteristic size of the resonance is  $r\sim1/p_0$, where $p_0=\sqrt{M E_0}$, and $p_0\approx 18.5$ MeV. Since $m_D \ll p_0$ when $T<1$ MeV, it is a good approximation to expand in $m_D r$ to lowest order and keep only the $r=0$ contribution. 
Despite $p_0\gg T$, the HTL approximation is still valid. The resonance property comes from the interplay between the contact strong and Coulomb interactions. The thermal correction of the Coulomb potential (both self energies and exchange) is controlled by the infrared scale $m_D$, which is much less than $T$. So the HTL approximation is valid when one considers the thermal correction on the Coulomb potential. Here, $p_0$ is the typical momentum transferred through the contact strong interaction rather than that of the Coulomb. The effect of the Coulomb interaction is felt over the entire path of the particles from infinity to nuclear contact, and thus not only sensitive to the momentum scale $p_0$.
Therefore
\be
\label{eq:green}
G(E,0,0;T\neq0) &=& \Big\langle {\bs 0} \Big|  \frac{1}{E-\hat{H}_0-\frac{Z^2\alpha}{r}+2 Z^2\alpha m_D +iZ^2\frac{32\alpha^2g(m_e\beta)T^3}{\pi m_D^2}  +i\epsilon}  \Big|  {\bs 0}  \Big\rangle\\
&=&G(\tilde{E},0,0;T=0) \,,
\ee
where
\be\label{shift}
\tilde{E} = E + 2 Z^2\alpha m_D +iZ^2\frac{32\alpha^2g(m_e\beta)T^3}{\pi m_D^2} \,.
\ee
Thus, in this approximation,  we  obtain the screened Coulomb Green's function in the plasma  by analytically continuing the vacuum Coulomb Green's function from $E$ to $\tilde{E}$. The function $C_{\eta}^2$ also needs to be analytically continued in the same way since the Coulomb wave function is the solution to an analogous analytic continuation of the Schr\"odinger equation. The scattering amplitude in the plasma can be written as
\be
T_{SC}&=&\frac{4\pi}{M}\frac{C_{\tilde{\eta}}^2e^{2i\sigma_0}}{-\frac{1}{a}+\frac{r_0}{2}p^2- \frac{P_0}{4}p^4-Z^2\alpha MH(\tilde{\eta})} \, ,
\ee
where $\tilde{\eta}$ is computed from $\tilde{E}$. Then the scattering amplitude squared is computed at different energies and fitted to the Breit-Wigner formula:
\be
\bigg(  \frac{4\pi}{M} \bigg)^2 \bigg| \frac{C_{\tilde{\eta}}^2e^{2i\sigma_0}}{-\frac{1}{a}+ \frac{r_0}{2}ME-\frac{P_0}{4} M^2E^2-Z^2\alpha MH(\tilde{\eta})}  \bigg|^2 =\frac{1}{p^2}\frac{A_0}{(E-E_r)^2+\Gamma^2/4}\,,
\ee
where $E_r$ is the resonance energy and $A_0$ is a constant.  An arbitrary constant $A_0$  appears in the numerator of our parametrization because the potential has an imaginary part which violates unitarity so the maximum amplitude is no longer the unitary limit. The total width $\Gamma$ is the sum of the thermal width, $\Gamma_{\ma{thermal}}$, caused by collisions with medium particles and the intrinsic width, $\Gamma_{\ma{intrinsic}}$, due to the spontaneous decay into two $\alpha$ particles.  $\Gamma_{\ma{intrinsic}}$ 
is defined as the width when only the static screening has been included, which has been calculated in Ref.~\cite{Yao:2016gny}. This contribution to the total width can be similarly calculated from the Coulomb 
Green's function with a real shift in the energy: $G(E,0,0;T\neq0) = G(E+Z^2\alpha m_D,0,0;T=0)$. 
\begin{figure}
\centering
\includegraphics[width=3in]{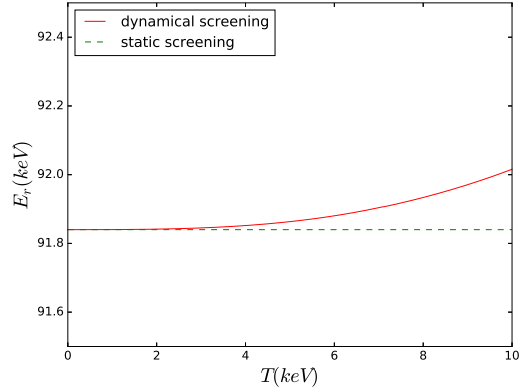}
\caption{The resonance energy, $E_r$, as a function of the temperature, $T$. The solid red line is the resonance energy with dynamical screening included and  the dotted green line is the energy when only static screening is included.}
\label{fig:resonance_energy}
\end{figure}
\begin{figure}
\centering
\includegraphics[width=3in]{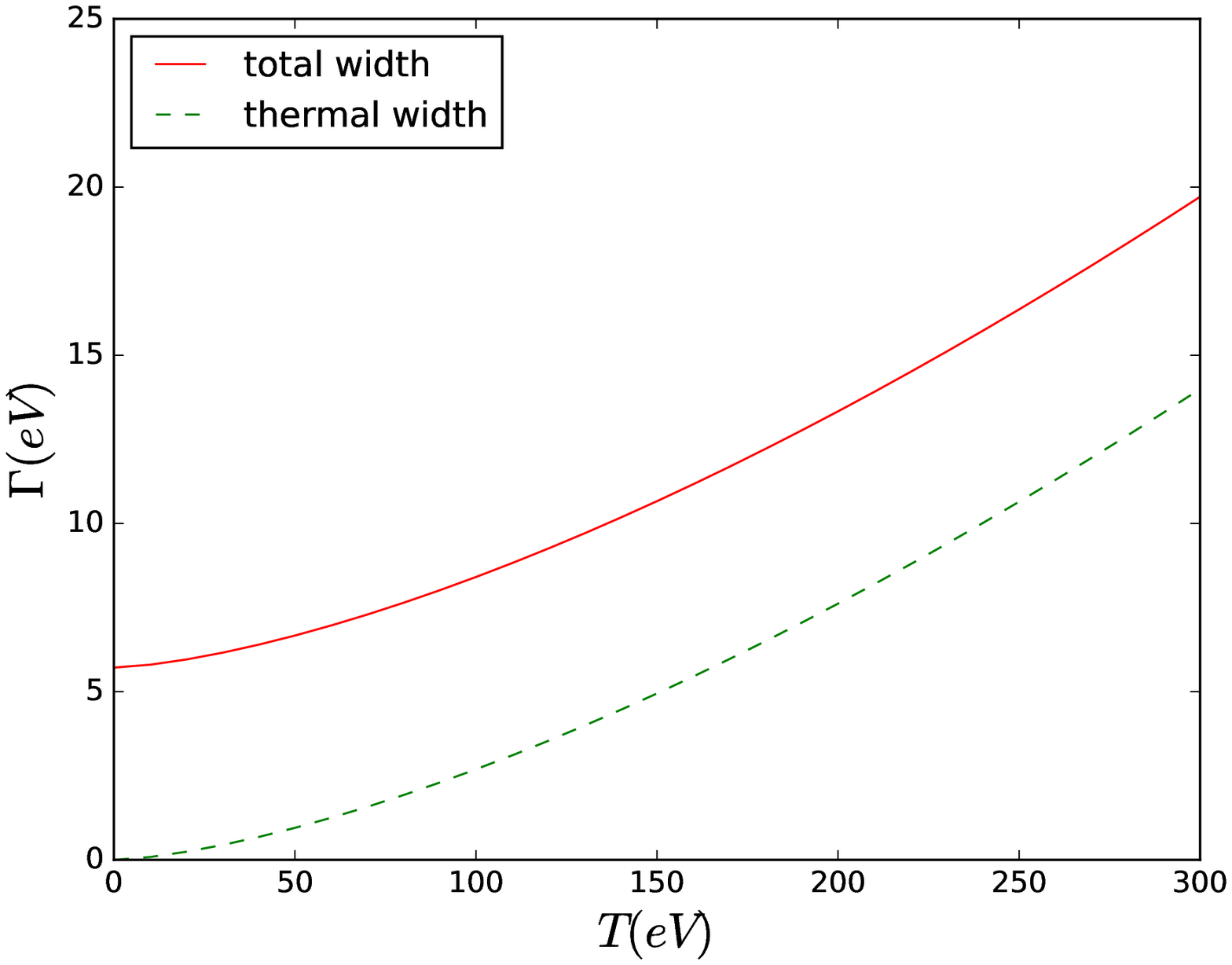}
\includegraphics[width=3in]{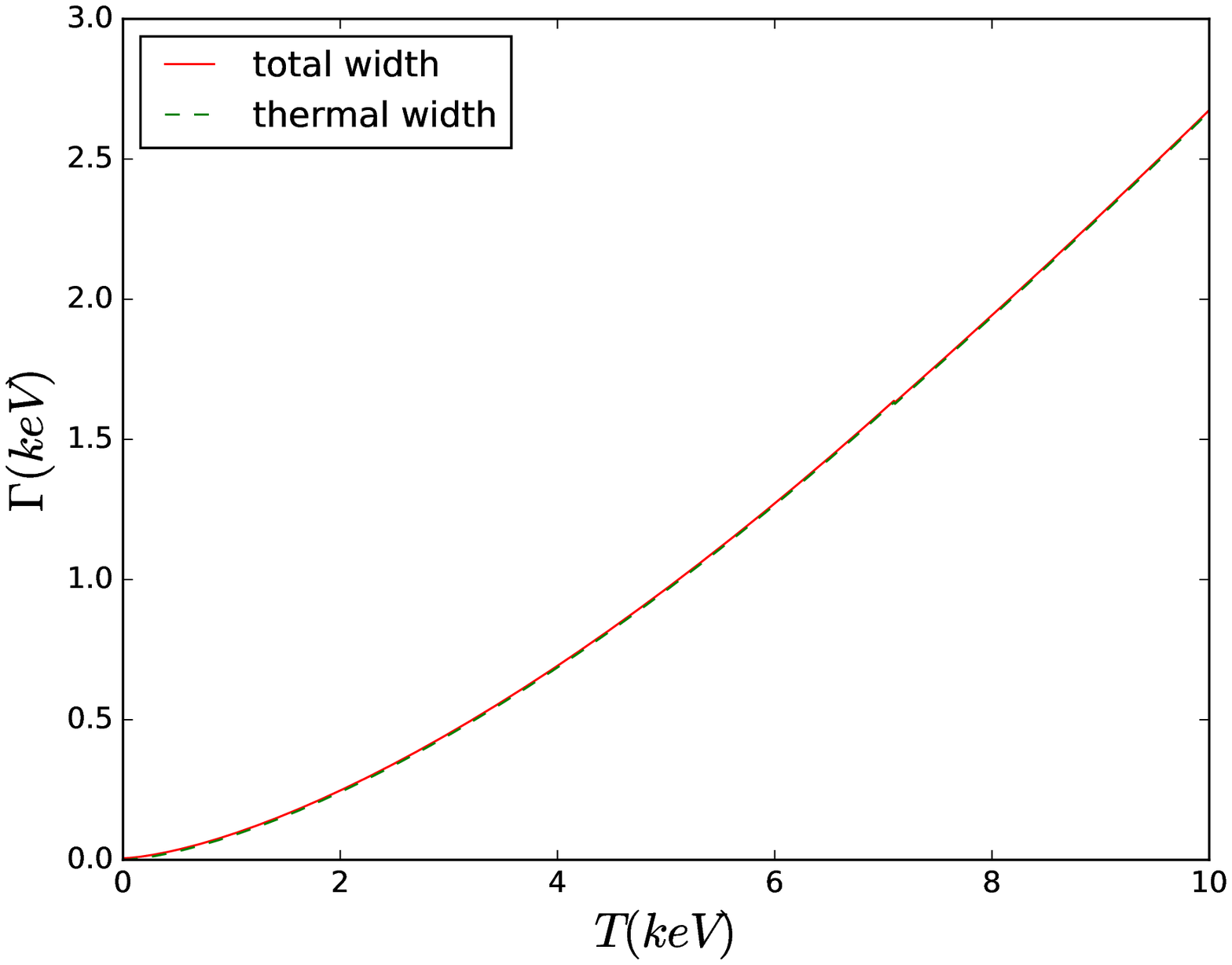}
\caption{The total width (red solid line) and the thermal width (green dotted line) as a function of plasma temperature, $T$, for $0 < T < 300$ eV (left) and $0 < T < 10$ keV (right).
}\label{fig:width}
\end{figure}

In Fig.~\ref{fig:resonance_energy} we plot the resonance energy up to $T=10$ keV since in this temperature range the resonance is well described by a Breit-Wigner form. 
The resonance energy is the red line and the green dotted line shows the resonance energy when only static screening effects are included. The resonance energy increases with plasma temperature due to the dynamical screening effect. When only static screening is included, the resonance energy decreases with temperature, but only very slightly in the temperature range shown. In Fig.~\ref{fig:width} the solid red line shows the total width of the resonance as a function of temperature and the green dotted line shows the thermal width. 
The total width is an increasing function of the temperature and for temperatures of $\ml{O}({\rm keV})$ the width is dominated by the thermal width due to dynamical screening.

That the resonance energy increases with the plasma temperature after taking into account the dynamical screening can be understood as follows: The imaginary potential describes the probability loss in the elastic channel as the two $\alpha$ particles move toward each other, which leads to the suppression of the wave function. This suppression is similar to that caused by a repulsive real potential. The imaginary potential and the associated suppression effect increase with the plasma temperature. As a result, it requires a higher kinetic energy to bring the two $\alpha$ particles into the nuclear interaction range. Therefore the resonance energy increases. The resonance also becomes wider. This effect obviously vanishes when $T=0$ and increases with the plasma temperature. Its value is comparable to the intrinsic width $\Gamma_0 = 5.57$ eV when $T \approx 160$ eV, and for temperatures of $\ml{O}$(keV) the thermal width completely dominates the total width of the resonance.

Although the resonance line shape is no longer well fitted by a Breit-Wigner above T = 10 keV, we can still calculate the scattering amplitude and thus the cross section accurately at any temperature. In the next section we calculate the thermally averaged $\alpha$-$\alpha$ nuclear scattering rate up to a temperature of  200 keV.

\section{Thermal Nuclear Scattering Rate}
\label{sect:rate}
The thermal nuclear scattering (or reaction) rate is defined as the thermally averaged product of the cross section and the relative velocity.
\be\label{thermal}
\langle \sigma v\rangle_T = \frac{\int\frac{\diff^3p}{(2\pi)^3}\sigma(p, T)\frac{p}{M/2}e^{-p^2/MT}}{\int\frac{\diff^3p}{(2\pi)^3}e^{-p^2/MT}}= \frac{4}{\sqrt{\pi M}}\frac{1}{T^{3/2}}\int \diff E E\sigma(E, T)e^{-E/T}.
\ee
Traditional calculations of the reaction rate use the cross section in vacuum, i.e., $\sigma(E, T)=\sigma(E, 0)$ or just include the static screening by shifting the resonance energy, which has little effect on the reaction rate. But now we have a way to estimate $\sigma(E, T\neq0)$ including both static and dynamical screening. As an example, we study the dynamical screening effect on the $\alpha$-$\alpha$ thermal nuclear scattering rate. We use the term ``scattering rate" since the process is elastic scattering of $\alpha$ particles, not a nuclear reaction. In the calculation of the cross section, we only use the scattering amplitude from the Coulomb modified nuclear interaction, $T_{SC}$, since the pure Coulomb contribution to the low-energy scattering amplitude is not expected to be important, especially in the resonance region.

\begin{figure}
\centering
\includegraphics[width=3.0in]{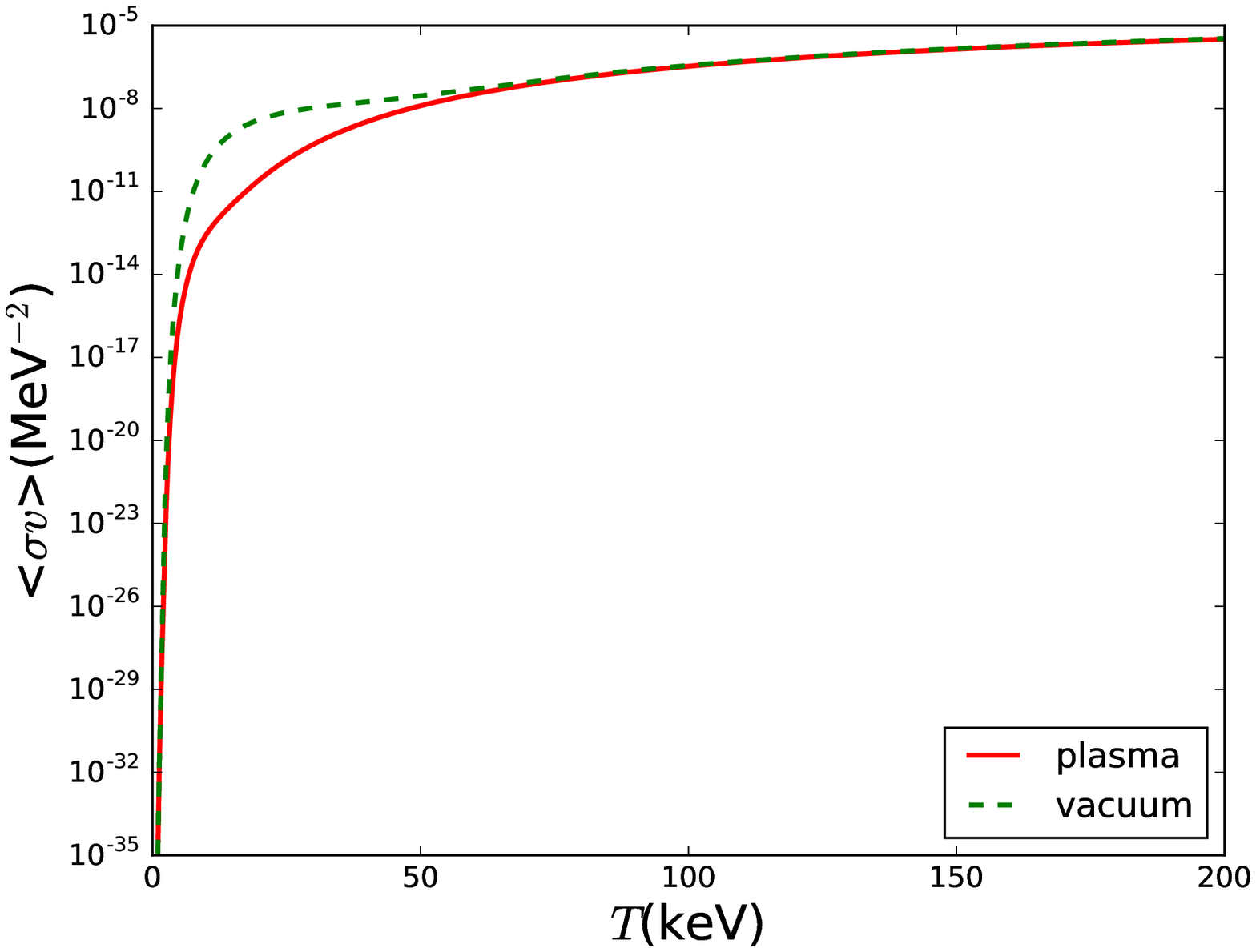}
\includegraphics[width=3.0in]{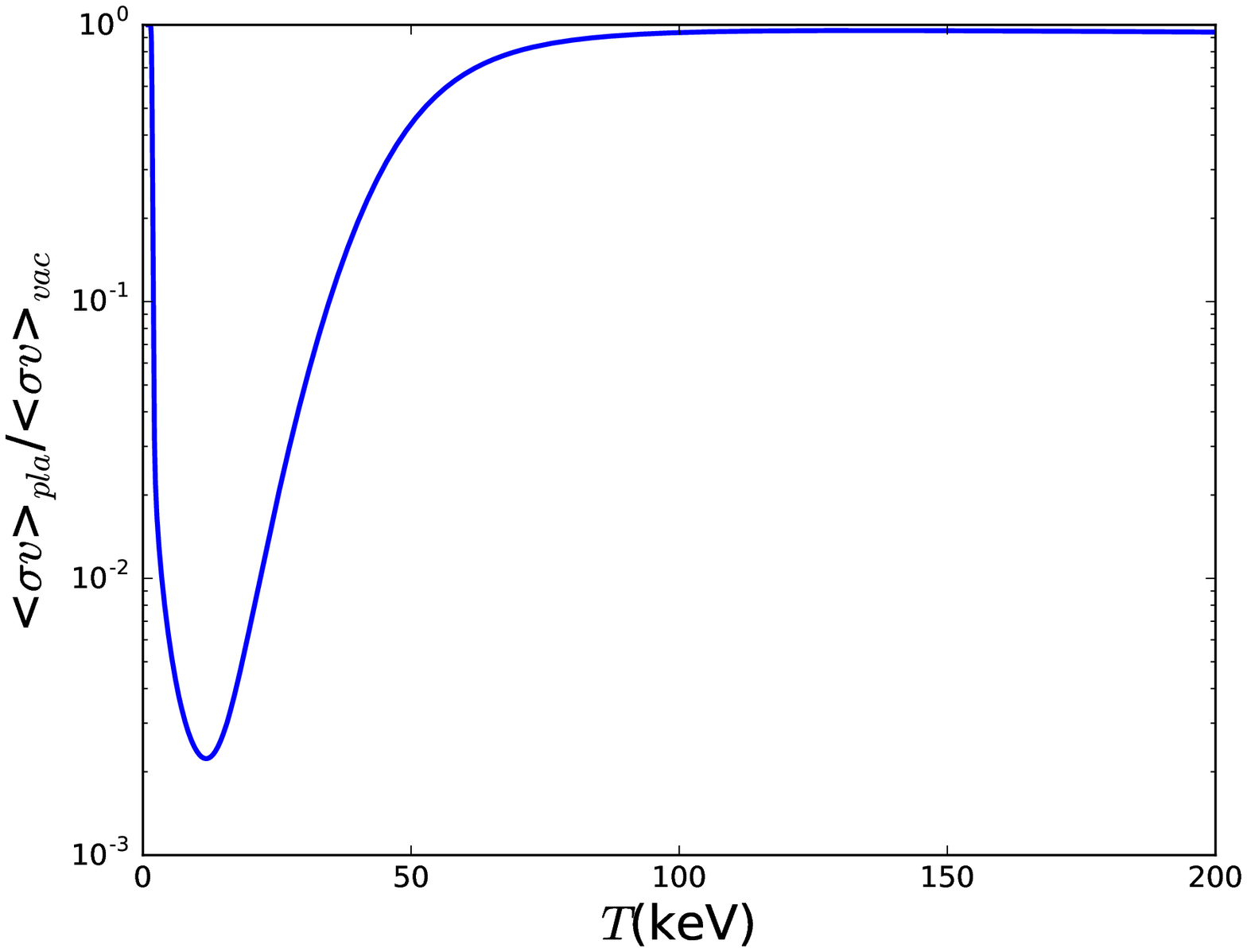}
\caption{Thermal $\alpha$-$\alpha$ nuclear scattering rate computed with $\sigma(E, T\neq0)$ (red solid line) and that with $\sigma(E, T=0)$ (green dotted line) as a function of the temperature (left). The ratio of the two thermal scattering rates as a function of temperature (right).}
\label{fig:rate_ratio}
\end{figure}

When computing the thermal $\alpha$-$\alpha$ scattering rate in the temperature range $T\in[1, 200]$ keV, we numerically integrate $E$ from $10$ keV to $3$ MeV in Eq.~(\ref{thermal}). The cross section is almost vanishing when $E<10$ keV since the phase shift there is almost zero. The energy region beyond $3$ MeV is omitted because for the temperatures we consider, the exponential factor $e^{-E/T}$ is extremely small. Physically, this means that at low temperatures, high energy cross sections contribute little to the scattering rate because the probability of having such high energy particles is exponentially suppressed. The calculated thermal scattering  rate with $\sigma(E, T\neq0)$, including dynamical screening,  and that with $\sigma(E, T=0)$ are shown in the left panel of Fig.~\ref{fig:rate_ratio} as the red solid and green dotted lines, respectively. The ratio of these thermal rates is shown in the right panel of Fig.~\ref{fig:rate_ratio}. The ratio approaches unity as the temperature goes to zero and also at high temperatures. The behavior in either limit is easy to understand. As the temperature goes to zero, the thermally averaged rate should be dominated by the threshold cross section which is below the resonance and is very 
small in either vacuum or plasma case. For high temperatures, the thermal scattering rate is dominated by high energy scattering. For high energies of $\ml{O}(100 \, {\rm keV})$, the complex shift in 
$\tilde E$ in Eq.~(29) is very small compared to $E$, so the effect of screening on the cross section is negligible. The ratio is far away from unity in the temperature range 1--75 keV, with the maximum suppression occurring at $T \sim 10$ keV. At this temperature the thermal scattering rate is almost $900$ times smaller once the screening effects are included. 
Our results indicate that dynamical screening effects have the greatest impact on the thermal scattering rate at temperatures just below the vacuum resonance energy. While here we demonstrate
this for $\alpha$-$\alpha$ scattering we expect this will also be the case for nuclear reactions that are dominated by above threshold resonances in the thermal domain.

\section{Conclusion}
\label{sect:concl}
We studied the QED plasma dynamical screening effect on the $\alpha$-$\alpha$ resonant scattering, where the resonant state is the $^8$Be nucleus. Collisions with medium particles result in an imaginary part of the $\alpha$-$\alpha$ potential, which leads to a thermal width of the resonance and the loss of unitarity in the scattering amplitude. Dynamical screening effects dominate over static screening effects and both the resonance energy and the width increase with the plasma temperature. Due to the loss of unitarity and increased width, the resonant cross section is highly suppressed and it is found that the $\alpha$-$\alpha$ thermal nuclear scattering rate is suppressed by a factor of $\sim 900$  when $T\sim10$ keV. Our calculations indicate that the dynamical screening effect in the plasma can be very large for nuclear reaction rates when those rates are dominated by above threshold resonances in the thermal domain. Such reactions can be important for nuclear astrophysics and cosmology. For example, a reaction that is critical in Big Bang Nucleosynthesis (BBN) predictions for the primordial $^7$Li abundance is $^7$Be(n, p)$^7$Li. This reaction is dominated by a resonant state of $^8$Be with $(J^P, I)=(2^-,0)$. The resonance exists approximately $0.01$ MeV above the $^7$Be+n threshold and thus lies well inside the thermal domain. A significant modification due to the dynamical screening on its reaction rate is expected and could have a significant impact on the prediction for the $^7$Li abundance, which is currently over-predicted by a factor of $\sim 3$, with a statistical significance of 4-5 
$\sigma$~\cite{Cyburt:2003fe,Cyburt:2015mya,Fields:2011zzb}. In future work, we plan to investigate the effect of dynamical screening on the $^7$Be(n, p)$^7$Li and other resonance dominated reactions that are important for cosmology and astrophysics.

\begin{acknowledgments}
X.Y. acknowledges the hospitality of the nuclear theory group at the Brookhaven National Laboratory where part of this work was completed. X.Y. also thanks the useful discussions with Yukinao Akamatsu and Krzysztof Redlich. T.M. acknowledges the hospitality of the particle theory group at UC-Irvine, and the organizers of the 2016 ESI Program on ``Challenges and Concepts for Field Theory and Application in the Era of the LHC Run-2".  B.M. and X.Y. are supported by U.S. Department of Energy Research Grant No. DE-FG02-05ER41367. T.M. is supported by U.S. Department of Energy Research Grant No. DE-FG02-05ER41368.
\end{acknowledgments}

\appendix
\section{Justification of HTL Photon Propagator}
\label{appenA}
In thermal field theory, the time-ordered propagator can be written in terms of the retarded and advanced propagators
\be
D_{\mu\nu} (q_0,\bs q)=\frac{1}{2}\big( D_{\mu\nu}^R (q_0,\bs q)+D_{\mu\nu}^A (q_0,\bs q)   \big) + \big(\frac{1}{2}+n_B(q_0)\big)\big( D_{\mu\nu}^R (q_0,\bs q)-D_{\mu\nu}^A (q_0,\bs q)   \big)\, .
\ee
For a photon propagator in a QED plasma, the form in the Coulomb gauge is
\be
D_{00}^{R(A)}(q_0,\bs q) &=& \frac{i}{\bs q^2-\Pi_{00}^{R(A)}(q_0,\bs q)}\,,
\ee
where the polarization tensor for $q_0\ll T,m_e,|\bs q|$ is given by
\be
\Pi_{00}^{R(A)}(q_0,\bs q)&=&-\frac{e^2}{\pi^2}\int_0^{\infty}\frac{p^2\diff p}{E_p}n_F(E_p)\bigg[  2+ \Big( \frac{|\bs q|}{2p}-\frac{2E_p^2}{p|\bs q|}   \Big)\ln\Big| \frac{|\bs q|-2p}{|\bs q|+2p} \Big|  \bigg] \\\nn
&&+(-) i\frac{q_0}{|\bs q|}\frac{2e^2}{\pi}\int_{|\bs q|/2}^{\infty}p \diff p \mathop{} n_F(E_p)\,.
\ee
Due to the non-relativistic feature of the system, $q_0$ can be set to be zero (see Eqs.~(\ref{eqn:sigma},\ref{eqn:coulomb})), so this is a valid approximation.
The Debye mass is defined as $m_D^2\equiv -\Re\Pi_{00}^{R(A)}(q_0=0,\bs q\rightarrow0)$. One can expand the retarded and advanced propagators as
\be
&&D_{00}^{R(A)}(q_0,\bs q) =  \frac{i}{(\bs q^2+m_D^2)\big(1 - \frac{m_D^2+\Re\Pi_{00}^{R(A)}(q_0,\bs q) + i\Im\Pi_{00}^{R(A)}(q_0,\bs q)}{\bs q^2+m_D^2}\big)} \\
&=& \frac{i}{\bs q^2+m_D^2} +(-) \frac{q_0}{|\bs q|(\bs q^2+m_D^2)^2}\frac{2e^2}{\pi}\int_{|\bs q|/2}^{\infty}p \diff p \mathop{} n_F(E_p)+\ml{O}\Big(q_0^2,\frac{m_D^2+\Re\Pi_{00}^{R(A)}(0,\bs q)}{\bs q^2+m_D^2}\Big)\,.
\ee
The expansion is valid because we will set $q_0=0$ in the end and $\frac{m_D^2+\Re\Pi_{00}^{R(A)}(0,\bs q)}{\bs q^2+m_D^2}$ is tiny for all values of $|\bs q|$. Then we have 
\be
D_{00}(q_0 = 0,\bs q) = \frac{i}{\bs q^2+m_D^2} + \frac{T}{|\bs q|(\bs q^2+m_D^2)^2}\frac{4e^2}{\pi}\int_{|\bs q|/2}^{\infty}p \diff p \mathop{} n_F(E_p)\,.
\ee

Since the dominant effect is the imaginary part of the potential, we will focus on the error in the imaginary potential caused by the HTL approximation.
The imaginary potential, with and without making the HTL approximation, is given by
\be
W(r) &=& -C\int\frac{\diff^3 q}{(2\pi)^3}(1+e^{i\bs q\cdot\bs r})\frac{T}{|\bs q|(\bs q^2+m_D^2)^2}\int_{|\bs q|/2}^{\infty} p\diff p\mathop{}n_F(E_p)\\
W^{\ma{HTL}}(r) &=& -C\int\frac{\diff^3 q}{(2\pi)^3}(1+e^{i\bs q\cdot\bs r})\frac{T}{|\bs q|(\bs q^2+m_D^2)^2}\int_{0}^{\infty} p\diff p\mathop{}n_F(E_p)\,,
\ee
where $C=64\alpha^2Z^2\pi$. Rescaling the integral variables: $|\bs q|=m_Dx$, $p\diff p =E_p\diff E_p$ and $E_p = Ty$, and doing the angular integral on $\bs q$, one obtains
\be
W(r) &=& -C'\frac{T^3}{m_D^2}\int_0^{\infty}\frac{x\diff x}{(1+x^2)^2}\Big(1+\frac{\sin(xm_Dr)}{xm_Dr}\Big)\int_{\frac{\sqrt{m_e^2+x^2m_D^2/4}}{T}}^{\infty}\frac{y\diff y}{e^y+1}\\
W^{\ma{HTL}}(r) &=& -C'\frac{T^3}{m_D^2}\int_0^{\infty}\frac{x\diff x}{(1+x^2)^2}\Big(1+\frac{\sin(xm_Dr)}{xm_Dr}\Big)\int_{\frac{m_e}{T}}^{\infty}\frac{y\diff y}{e^y+1}\,,
\ee
where $C'=\frac{32\alpha^2Z^2}{\pi}$. The error is defined as $\Delta W(r)\equiv  W(r)-W^{\ma{HTL}}(r)$:
\be
\Delta W(r) = C'\frac{T^3}{m_D^2}\int_0^{\infty}\frac{x\diff x}{(1+x^2)^2}\Big(1+\frac{\sin(xm_Dr)}{xm_Dr}\Big)\int^{\frac{\sqrt{m_e^2+x^2m_D^2/4}}{T}}_{\frac{m_e}{T}}\frac{y\diff y}{e^y+1}\,.
\ee
In the main text we show that the resonance size scales as $r\sim1/(18.5\ \ma{MeV})$ and for the temperature range considered $T\lesssim200$ keV, $m_Dr\ll1$ and one can set $r=0$. So
\be
W^{\ma{HTL}}(r=0) &=& -C'\frac{T^3}{m_D^2}\bigg[\frac{m_e}{T}\ln(1+e^{-m_e/T}) - Li_2(-e^{-m_e/T}) \bigg] \nn \\
&=& -C'\frac{T^3}{m_D^2}\Big(1+\frac{m_e}{T}\Big)e^{-m_e/T} + \ml{O}(e^{-2m_e/T} )\,.
\ee

Next we consider $\Delta W(r)$. The integrand is positive and $\frac{\sin{x}}{x}$ is decreasing with $x$, so we have
\be
\Delta W(r) &<& C'\frac{T^3}{m_D^2}\int_0^{\infty}\frac{2x\diff x}{(1+x^2)^2}\bigg[  -\frac{\sqrt{m_e^2+x^2m_D^2/4}}{T}\ln(1+e^{-\sqrt{m_e^2+x^2m_D^2/4}/T}) \\\nn
&&+ Li_2(-e^{-\sqrt{m_e^2+x^2m_D^2/4}/T}) + \frac{m_e}{T}\ln(1+e^{-m_e/T}) - Li_2(-e^{-m_e/T}) \bigg]\,.
\ee
Making a change of variable $z=x^2$ and defining $\tilde{m}_e=m_e/T$ and $\tilde{m}_D=m_D/T$, we find
\be
\Delta W(r) &<&C'\frac{T^3}{m_D^2}\int_0^{\infty}\frac{\diff z}{(1+z)^2}\bigg[  -\sqrt{\tilde{m}_e^2+z\tilde{m}_D^2/4}\ln(1+e^{-\sqrt{\tilde{m}_e^2+z\tilde{m}_D^2/4}}) \\\nn
&&+ Li_2(-e^{-\sqrt{\tilde{m}_e^2+z\tilde{m}_D^2/4}}) + \tilde{m}_e\ln(1+e^{-\tilde{m}_e}) - Li_2(-e^{-\tilde{m}_e}) \bigg]\,.
\ee
Integrating by parts $\frac{\diff z}{(1+z)^2}=-\diff \frac{1}{1+z}$ and noticing that the boundary terms vanish, we obtain
\be\nn
\Delta W(r) &<&  C'\frac{T^3}{m_D^2} \int_0^{\infty} \frac{1}{1+z}\diff\bigg[  -\sqrt{\tilde{m}_e^2+z\tilde{m}_D^2/4}\ln(1+e^{-\sqrt{\tilde{m}_e^2+z\tilde{m}_D^2/4}}) \\
&&+ Li_2(-e^{-\sqrt{\tilde{m}_e^2+z\tilde{m}_D^2/4}}) + \tilde{m}_e\ln(1+e^{-\tilde{m}_e}) - Li_2(-e^{-\tilde{m}_e}) \bigg]\,.
\ee
We then expand the logarithm and dilogarithm functions
\be\nn
\Delta W(r) &<&C'\frac{T^3}{m_D^2} \int_0^{\infty} \frac{\diff z}{1+z} \frac{\diff}{\diff z} \bigg[ -(\sqrt{\tilde{m}_e^2+z\tilde{m}_D^2/4}+1) e^{-\sqrt{\tilde{m}_e^2+z\tilde{m}_D^2/4}} \bigg] + \ml{O}(e^{-2m_e/T})\\
&=& C'\frac{T^3}{m_D^2} \int_0^{\infty} \frac{\diff z}{1+z}\frac{\tilde{m}_D^2}{8}e^{-\sqrt{\tilde{m}_e^2+z\tilde{m}_D^2/4}}+ \ml{O}(e^{-2m_e/T}) \\
&=& C'\frac{T^3}{m_D^2} e^{-m_e/T} \mathop{}\frac{\tilde{m}_D^2}{8}  \int_0^{\infty} \frac{\diff z}{1+z}e^{-\sqrt{1+z\frac{m_D^2}{4m_e^2}}}+ \ml{O}(e^{-2m_e/T})\,.
\ee
Taking the ratio gives
\be
\frac{\Delta W(r) }{|W^{\ma{HTL}}(0)|}\lesssim\frac{m_D^2}{8T(T+m_e)}\int_0^{\infty} \frac{\diff z}{1+z}e^{-\sqrt{1+z\frac{m_D^2}{4m_e^2}}}+ \ml{O}(e^{-m_e/T})\,.
\ee
Since $\lambda \equiv \frac{m_D^2}{4m_e^2}$ is small, the integral over $z$ can be approximated analytically. We want to evaluate
\be
I\equiv \int_0^{\infty}\frac{\diff z}{1+z} e^{-\sqrt{1+z\lambda}}\,.
\ee

Changing the variable from $z$ to $w=1+z\lambda$ leads to
\be
I = \int_1^{\infty} \frac{\diff w}{w-1+\lambda}e^{-\sqrt{w}}\,.
\ee
We cannot set $\lambda=0$ so far because it leads to a divergent integral, but we can do an integration by parts:
\be
I &=& \int_1^{\infty}e^{-\sqrt{w}}  \diff\ln(w-1+\lambda)\\
&=& e^{-\sqrt{w}}\ln(w-1+\lambda)\Big|_1^{\infty} - \int_1^{\infty} \ln(w-1+\lambda)e^{-\sqrt{w}} \frac{\diff w}{-2\sqrt{w}}\\
&=& -\frac{\ln\lambda}{e} + \int_1^{\infty} \ln(w-1)e^{-\sqrt{w}} \frac{\diff w}{2\sqrt{w}} + \ml{O}(\lambda)\\
&=& \frac{\ln(\frac{1}{\lambda})}{e}+0.176 + \ml{O}(\lambda)
\ee
Finally we have
\be
\frac{\Delta W(r) }{|W^{\ma{HTL}}(0)|}\lesssim\frac{m_D^2}{8T(T+m_e)}\Big( \frac{2}{e} \ln{\frac{2m_e}{m_D}} +0.176 \Big)+ \ml{O}\Big(\frac{m_D^2}{m_e^2},\ e^{-m_e/T}\Big)\,.
\ee
We find that the error introduced by the HTL approximation is suppressed by $\frac{m_D^2}{Tm_e}\ln{\frac{m_e}{m_D}}$ when $T$ is small ($T\lesssim200$ keV). 

Physically, the resonance property is determined by both the contact strong and the Coulomb interactions. The Coulomb interaction is long-ranged and thus sensitive to low-energy scales. The scale of the screened Coulomb interaction is set by the Debye mass, which is much smaller than the temperature. When studying thermal corrections on the Coulomb potential, one can make use of the HTL approximation. The fact that the resonance momentum $p_0$ is much larger than the Debye mass indicates that one can set $r=0$ for the thermal correction of the potential in the calculation of the Green's function.

\end{document}